**Resonant plane waves in metamaterials with dipoles and quadrupoles coupled with quantum system**

A. Chipouline, M. Dobynde, T. Pertsch

Contents

## 1. Introduction

The mitigation of the optical losses in metamaterials (MMs) could be potentially achieved by a combination of lower loss materials [1], [2] and by providing gain by doping of the MMs with optically active emitters – see recent reviews [3], [4]. It has been shown experimentally that this form of loss compensation does not prohibit the negative index property of the MM [5]. In addition, the coupling with optically active emitters can compensate losses of the plasmonic components making them feasible for telecom applications [6], [7], [8], [9]. Along with the plasmonic waveguides, other active components like modulators and switchers form a full-scale nomenclature for application in the next generation signal processing devices [10]. Several types of theoretical models have been suggested in order to describe gain processes in MMs and plasmonic waveguides. Analytical or semi-analytical models [11], [12] used the density matrix approach for the quantum dynamics description from the very beginning, but to the best of our knowledge have not been combined with the multipole approach [13] and consequently do not provide an adequate enough platform to investigate the properties of MMs fully. Instead, the vast majority of the publications have utilized the computational methods [14], [4] which give results close to the experimentally realizable data. Unfortunately, in some cases the numerical approach cannot subdivide different physical effects and as a result hides or limits any physical insight. For example, for the case of competition between spaser eigen generation and dynamics driven by an external field the numerical approaches mix all fields (generated by MAs and the external one) in one and thus far cannot demonstrate instability effect in the form given by analytical analysis [15].

Here the multipole approach [13], in combination with the density matrix formalism is used for establishing of the model for MMs with gain. This approach allows us to investigate analytically or semi-analytically the interplay between gain and magnetic properties of the MMs, the influence of internally unstable operation mode for spasers (MAs coupled with emitters, which the MMs consist of) on the propagation characteristics, and finally to optimize MM design. Moreover, the presented model is in line with the previously presented approach [16] (actually is its natural extension on the problem of plane wave propagation) and from the other side pretty clear and observable, which makes the model a perfect platform for various university courses.

## 2. Model of dipole coupled with quantum system (QS)

In this chapter, an analytical model for describing complex dynamics of a hybrid system consisting of resonantly coupled classical resonator (classic system-CS) and quantum structures (quantum system-QS) is presented. CS in this model correspond to plasmonic nanoresonators of various geometries, as well as other types of nano- and microstructures, optical response of which can be described classically. QSs are represented by atoms or molecules, or their aggregates (for example, quantum dots, carbon nanotubes, dye molecules, polymer or bio molecules etc.), which can be accurately modeled only with the use of the quantum-mechanical approach. Our model is based on the set of equations that combines well-established density matrix formalism appropriate for quantum systems, coupled with harmonic-oscillator equations ideal for modeling sub-wavelength plasmonic resonators. Using the developed approach, regular and stochastic dynamics of the nanolaser (spaser) has been considered, and generalization of Schawlow-Towns expression has been elaborated [17]. The dynamics of the plane wave propagating in the MMs (where the metaatoms – MAs - are the coupled CS and QS) is considered in this paper.

Accurate description of the dynamics of resonantly interacting CSs is a fundamental problem. The current approach is to use a set of coupled equations for two (or more) harmonic oscillators, which can normally be solved under appropriate approximations. It combines mathematical simplicity with adequate physical insight and has been adopted in various branches of science ranging from optics to nuclear physics. If the interacting systems are QSs their dynamics can be satisfactory described in the framework of quantum mechanics based on the Schrödinger equation or density matrix approach, for instance. However, for describing CS and QS coupled together a special approach is required. It was originally developed to model the dynamics of lasers where the classical system is normally represented by an optical (mirror) resonator, while the quantum system – by amplifying medium [18]. The basic idea was that the quantum formalism allowed accurate calculation of the medium's polarizability, while the latter could be used in the classical Maxwell equations describing electromagnetic fields in the optical resonator.

With the rapid development of nanotechnology it has become possible to engineer and study hybrid CS&QS systems at the nanometer scale such as metallic nanoresonators and their arrays combined with quantum dots, carbon nanotubes or dye molecules [19], [20], [21], [22]. While optical response of a metallic nanoresonator is affected by plasmonic excitations and shape, its rather complicated dynamics can still be satisfactory modeled by the harmonic oscillator equations with appropriately chosen parameters [13]. This allows us to extend the quantum-classical treatment to modeling analytically a wide range of optical and plasmonic effects in the hybrid quantum MMs, such as loss compensation, enhancement of nonlinear response and luminescence, etc. Furthermore, the model can be used to describe the dynamics of superconducting Josephson-junction-based MMs, as well as SQUIDs coupled to an RF strip resonator [20].

Although a wide range of numerical approaches describing rigorously the internal quantum dynamics of molecules have been developed, including time dependent density function theory, multi-configurational self-consistent field method, polarizable quantum mechanical/molecular mechanical method, capacitance-polarizability interaction model and discrete interaction model/quantum mechanics (see [23], [24], [25], [26], [27] and references therein), our model takes advantage of the phenomenological approach, which allows relatively simple analytical treatment and provides deeper insight into the physical behavior of coupled CS&QS. The parameters of such a model can always be found from fitting experimental data and/or using rigorous numerical approaches mentioned above. Another advantage of the proposed model is that it takes into account additional important nonradiative relaxation channels due to stochastic interaction with the environment, which is naturally included in the adopted density matrix approach through two phenomenological relaxation times for polarization and population.

Here a QS placed in the near-field zone of a CS is considered – see Fig. 1. The field produced by the CS, $E_{CS}$, affects QS that in turn acts on CS with its field $E_{QS}$. In addition, there is an external field $E_{ext}$ of the incident light, which interacts with both CS and QS; see Fig. 1. The actual number of the harmonic-oscillator equations required to adequately describe CS depends on its particular structure [13]. For the illustration purpose the analysis will be first restricted to just one dipole like MA (Fig. 2(a)), which is described by a single harmonic-oscillator equation), and then extended to the quadrupole-like based CS possessing a magnetic response – Fig. 2(b). The dynamics of QS is modelled using the density matrix formalism.

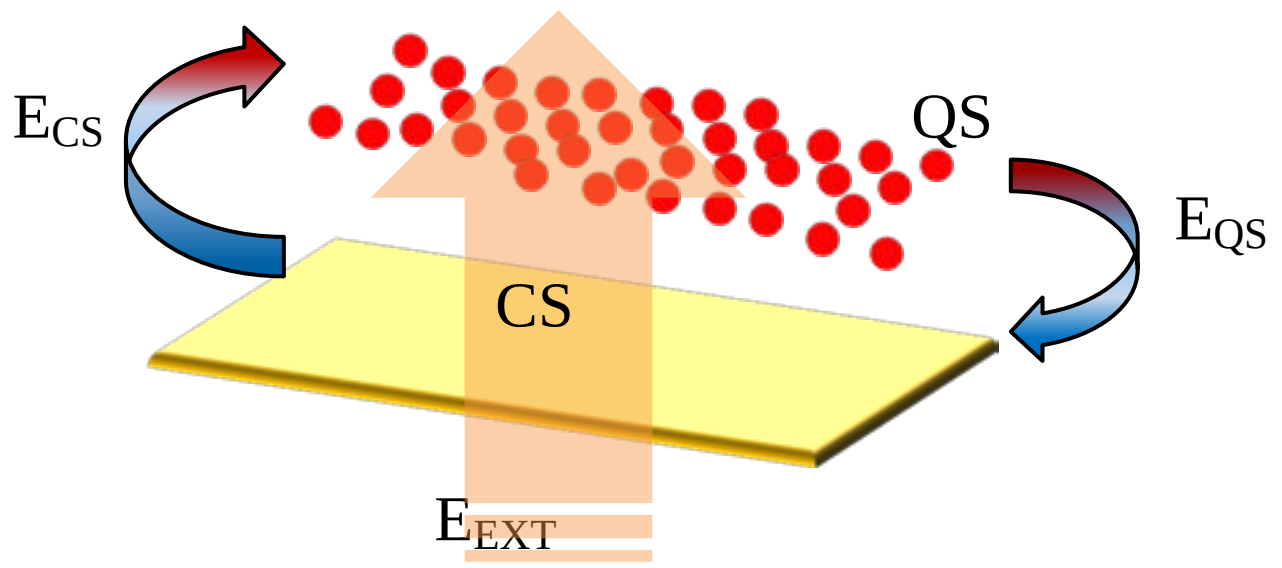

**Fig. 1: Schematic representation of the interaction between plasmonic nanoresonator (Classic System – CS, yellow block) covered with a layer of Quantum Systems (Quantum System – QS, red circles). $E_{CS}$ is the field produced by CS and acting on QS, $E_{QS}$ is the field produced by QS and acting on CS, $E_{EXT}$ is the external filed field acting on both CS and QS.**

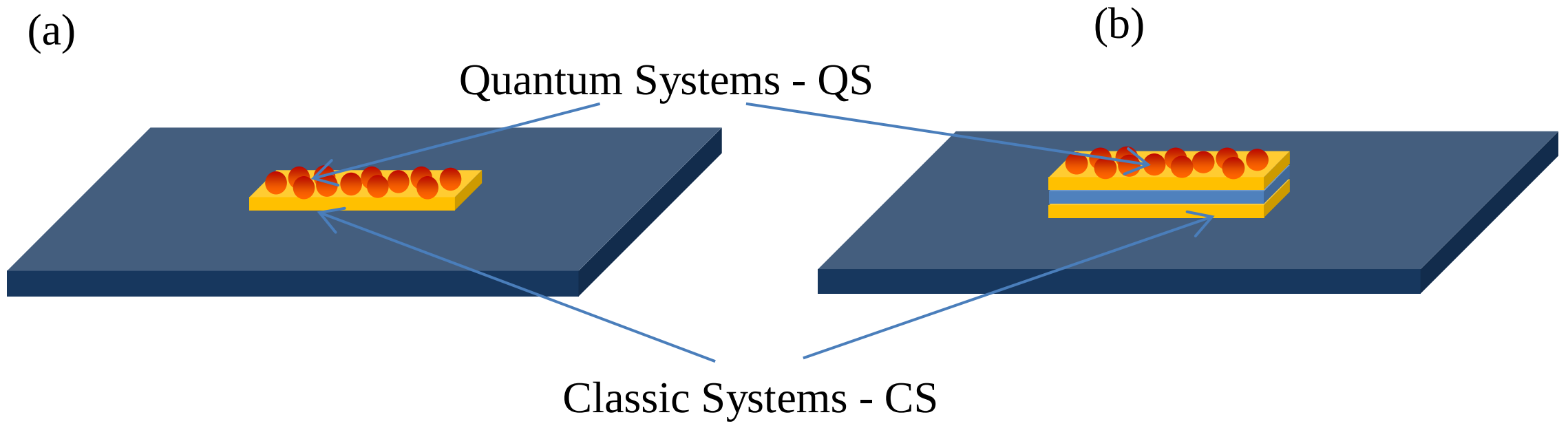

**Fig. 2: Schematic of the modelled active hybrid MAs with quantum ingredients: a plasmonic nanoresonator (Classic System – CS, yellow blocks) covered with a layer of quantum ingredients (Quantum System – QS, red circles). (a) – Dipole-like MA (one nanoresonator, (b) – Quadrupole-like MA (two coupled nanoresonators, separated by dielectric layer).**

In general case the quantum dynamics of QS that is assumed to be in contact with a thermostat environment, is described by the following set of ordinary differential equations [28], [29]:

$$\begin{cases} \dfrac{d\rho_{nn}}{dt} + \sum_m \left(k_{nm}\rho_{nn} - k_{mn}\rho_{mm}\right) = -\dfrac{i}{\hbar}\sum_m \left(H_{nm}\rho_{mn} - H_{mn}\rho_{nm}\right) \\ \dfrac{d\rho_{kl}}{dt} + i\omega_{kl}\rho_{kl} + \dfrac{\rho_{kl}}{\tau_{kl}} = -\dfrac{i}{\hbar}\sum_m \left(H_{km}\rho_{ml} - H_{ml}\rho_{km}\right) \end{cases} \quad (1)$$

Here $k_{nm}$ and $\tau_{kl}$ are energy and phase relaxation constants respectively, $\omega_{kl}$ is frequency of the transition from $k$ to $l$, $\rho_{nn}$ and $\rho_{kl}$ ($\rho_{kl} = \rho^*_{kl}$) are diagonal and non-diagonal elements of the density matrix, $H_{kl}$ is a Hamiltonian matrix element responsible for interaction of the quantum system with the external field. In framework of this formalism the averaged polarization density is expressed through non-diagonal density matrix elements:

$$P_{kl} = N\mu_{kl}\left(\rho_{kl} + \rho_{lk}\right) \tag{2}$$

$\mu_{kl}$ is the dipole moment of a quantum system, which is proportional to overlap integral between psi-functions of both levels, $N$ is the quantum systems concentration.

In the case of resonant interaction the internal QS dynamics can be modelled to a first approximation by a two-level system subjected to a pump:

$$\begin{cases} \dfrac{d\rho_{12}}{dt} - i\omega_{21}\rho_{12} + \dfrac{\rho_{12}}{\tau_2} = -\dfrac{iH_{12}\left(\rho_{22}-\rho_{11}\right)}{\hbar} \\ \dfrac{d\rho_{22}}{dt} + \dfrac{\rho_{22}}{\tilde{\tau}_1} = -\dfrac{iH_{12}\left(\rho_{12}-\rho_{21}\right)}{\hbar} + W\rho_{11} \\ \rho_{22} + \rho_{11} = 1 \end{cases} \tag{3}$$

Here $\rho_{22}$, $\rho_{11}$ and $\rho_{12}$, $\rho^*_{12}$ are the diagonal and non-diagonal matrix density elements, respectively; $\tau_2$ and $\tilde{\tau}_1$ are the constants describing phase and energy relaxation processes due to the interaction with a thermostat; $\omega_{21} = (E_2 - E_1)/\hbar$ is the transition frequency between levels 2 and 1; $H_{12}$ is the Hamiltonian matrix element responsible for interaction of QS with the external fields; $W$ is the phenomenological pump rate – this could model pumping QS. It is also convenient to introduce new variables $N = \rho_{22} - \rho_{11}$ and $N_0 = \dfrac{\left(W\tilde{\tau}_1 - 1\right)}{\left(W\tilde{\tau}_1 + 1\right)}$ so that:

$$\begin{cases} \dfrac{d\rho_{12}}{dt} + i\omega_{12}\rho_{12} + \dfrac{\rho_{12}}{\tau_2} = -\dfrac{iH_{12}N}{\hbar} \\ \dfrac{dN}{dt} + \dfrac{N - N_0}{\tau_1} = -\dfrac{2iH_{12}\left(\rho_{12}-\rho^*_{12}\right)}{\hbar} \\ \tau_1 = \dfrac{\tilde{\tau}_1}{W\tilde{\tau}_1 + 1} \end{cases} \tag{4}$$

In order to describe dynamics of the plasmonic nanoresonator the following harmonic-oscillator equation is used:

$$\frac{d^2x}{dt^2}+2\gamma\frac{dx}{dt}+\omega_0^2 x=\chi\left(\mathrm{E}_{ext}+\mathrm{E}_{QS}\right) \tag{5}$$

Here $\gamma$ and $\omega_0$ are the loss coefficient and resonance eigenfrequency, $E_{\text{ext}}$ and $E_{\text{QS}}$ are the external electric field and field generated by QS respectively, and $\chi$ is the effective kinetic inductance of the nanoresonator. The dimensionless variable $x$ corresponds here to one the dynamic characteristics of the oscillator, which will be identified later.

From (4) and (5) one can obtain:

$$\begin{cases} \dfrac{d\rho_{12}}{dt}+i\omega_{12}\rho_{12}+\dfrac{\rho_{12}}{\tau_2}=-\dfrac{iH_{12}N}{\hbar} \\ \dfrac{dN}{dt}+\dfrac{N-N_0}{\tau_1}=-\dfrac{2iH_{12}\left(\rho_{12}-\rho_{12}^*\right)}{\hbar} \\ \dfrac{d^2x}{dt^2}+2\gamma\dfrac{dx}{dt}+\omega_0^2 x=\chi\left(\mathrm{E}_{ext}+\mathrm{E}_{QS}\right) \end{cases} \tag{6}$$

In order to make the next step it is necessary to determine the nature of the interaction between CS and QS and write down expressions for $H_{12}$ and $E_{\text{QS}}$. We assume that the fields in the near-field zone of both systems are predominantly electric and produced by the effective electric dipole moments *d*. Electric field of an oscillating dipole is proportional to the *d*:

$$E\sim d \tag{7}$$

Correspondingly, electric field generated by the dipole moment of QS $d_{QS}$ at the location of CS can be written as:

$$E_{QS}\sim d_{QS}\sim\mu_{QS}\left(\rho_{12}+\rho_{21}\right) \tag{8}$$

where $\mu_{\text{QS}}$ is the dipole moment of QS.

According to the same relation the local electric field of CS is:

$$E_{CS}\sim d_{CS}\sim\mu_{CS}x \tag{9}$$

where $\mu_{CS}$ is the effective dipole moment of CS. From (9) it follows that the dimensionless variable $x$ has basically the same meaning as the non-diagonal element of the density matrix, namely the dimensionless polarization. It is worth noting that (8) and (9) assume both QS and CS as point-like dipoles. Despite the evident importance of addressing the overlap between the spatially inhomogeneous field of the plasmonic nanoresonator and localization of the quantum system, it is believed that this corresponds to the next level of complication that is not essential for adequate modeling of the response dynamics. The Hamiltonian of interaction $H_{12}$ is defined by the following expressions:

$$\begin{cases} H_{12} = -\mu_{QS}\left(\mathrm{E}_{ext} + E_{CS}\right) = -\left(\mu_{QS}\,\mathrm{E}_{ext} + \alpha_x x\right) \\ \alpha_x \sim \mu_{QS}\mu_{CS} \end{cases} \tag{10}$$

Substituting (8) and (10) into (6) one obtains:

$$\begin{cases} \dfrac{d\rho_{12}}{dt} - i\omega_{21}\rho_{12} + \dfrac{\rho_{12}}{\tau_2} = \dfrac{i\left(\mu_{QS}\,\mathrm{E}_{ext} + \alpha_x x\right)N}{\hbar} \\ \dfrac{dN}{dt} + \dfrac{N - N_0}{\tau_1} = \dfrac{2i\left(\mu_{QS}\,\mathrm{E}_{ext} + \alpha_x x\right)\left(\rho_{12} - \rho_{21}\right)}{\hbar} \\ \dfrac{d^2x}{dt^2} + 2\gamma\dfrac{dx}{dt} + \omega_0^2 x - \alpha_\rho\left(\rho_{12} + \rho_{21}\right) - \chi\,\mathrm{E}_{ext} = 0 \\ N_0 = \dfrac{\left(W\tilde{\tau}_1 - 1\right)}{\left(W\tilde{\tau}_1 + 1\right)}, \quad \tau_1 = \dfrac{\tilde{\tau}_1}{W\tilde{\tau}_1 + 1} \\ \alpha_\rho \sim \mu_{QS}\chi \\ \alpha_x \sim \mu_{QS}\mu_{CS} \end{cases} \tag{11}$$

Here $N_0$ is the population inversion due to pump (in the absence of pump $N_0 = -1$); $N_0 > 0$ corresponds to the regime of amplification, $N_0 < 0$ – to losses. Both eigenfrequencies $\omega_{21}$ and $\omega_0$ are the resonance frequencies of QS and CS respectively and can vary independently. Rotating wave approximation for the system (11) is introduced through the following notations:

$$\begin{cases} \rho_{12} = \dfrac{\tilde{\rho}_{12}}{2}\exp(i\omega t) \\ x = \dfrac{1}{2}\left(\tilde{x}(t)\exp\left(-i\omega t\right) + \tilde{x}(t)^*\exp\left(i\omega t\right)\right) \\ \mathrm{E}_{ext} = \dfrac{1}{2}\left(A(t)\exp\left(-i\omega t\right) + A(t)^*\exp\left(i\omega t\right)\right) \end{cases} \tag{12}$$

resulting in:

$$\begin{cases} \frac{d\tilde{\rho}_{12}}{dt}+\tilde{\rho}_{12}\left(\frac{1}{\tau_2}+i\left(\omega-\omega_{21}\right)\right)=\frac{i\alpha_x\tilde{x}^*N}{\hbar}+\frac{i\mu_{QS}A^*N}{\hbar} \\ \frac{dN}{dt}+\frac{\left(N-N_0\right)}{\tau_1}=\frac{i\alpha_x\left(\tilde{x}\tilde{\rho}_{12}-\tilde{x}^*\tilde{\rho}_{12}^*\right)+i\mu_{QS}\left(A\tilde{\rho}_{12}-A^*\tilde{\rho}_{12}^*\right)}{2\hbar} \\ 2\left(\gamma-i\omega\right)\frac{d\tilde{x}}{dt}+\left(\omega_0^2-\omega^2-2i\omega\gamma\right)\tilde{x}=\alpha_\rho\tilde{\rho}_{12}^*+\chi A \end{cases} \tag{13}$$

(13) are the master set of equations describing regular dynamics of interacting QS and CS. Taking into account stochastic noise sources, the set becomes:

$$\begin{cases} \frac{d\tilde{\rho}_{12}}{dt}+\tilde{\rho}_{12}\left(\frac{1}{\tau_2}+i\left(\omega-\omega_{21}\right)\right)=\frac{i\alpha_x\tilde{x}^*N}{\hbar}+\frac{i\mu_{QS}A^*N}{\hbar}+\xi_\rho \\ \frac{dN}{dt}+\frac{\left(N-N_0\right)}{\tau_1}=\frac{i\alpha_x\left(\tilde{x}\tilde{\rho}_{12}-\tilde{x}^*\tilde{\rho}_{12}^*\right)+i\mu_{QS}\left(A\tilde{\rho}_{12}-A^*\tilde{\rho}_{12}^*\right)}{2\hbar} \\ 2\left(\gamma-i\omega\right)\frac{d\tilde{x}}{dt}+\left(\omega_0^2-\omega^2-2i\omega\gamma\right)\tilde{x}=\alpha_\rho\tilde{\rho}_{12}^*+\chi A+\xi_x \end{cases} \tag{14}$$

Here $\xi_\rho$ and $\xi x$ are the stochastic Langevin terms, which take into account spontaneous emission and thermal fluctuations respectively (the stochastic term influence description can be found in [17]).

## 3. Evaluation of numerical values of parameters

The phenomenological parameters in (14) can be evaluated based on the measurement data or by the fitting with the rigorous numerical calculations [13], [33]. For the numerical values of the plasmonic nanoresonator itself (without coupling with the quantum emitters) the eigen frequency and damping coefficient can be easily taken from the measurements (for example, spectrum of losses) or numerical data – the eigen frequency $\omega_0$ corresponds to the maximum losses and the damping coefficient $\gamma$ is recalculated from the bandwidth. As like in [13] we assume here the structure with the sizes depicted in Fig. 3:

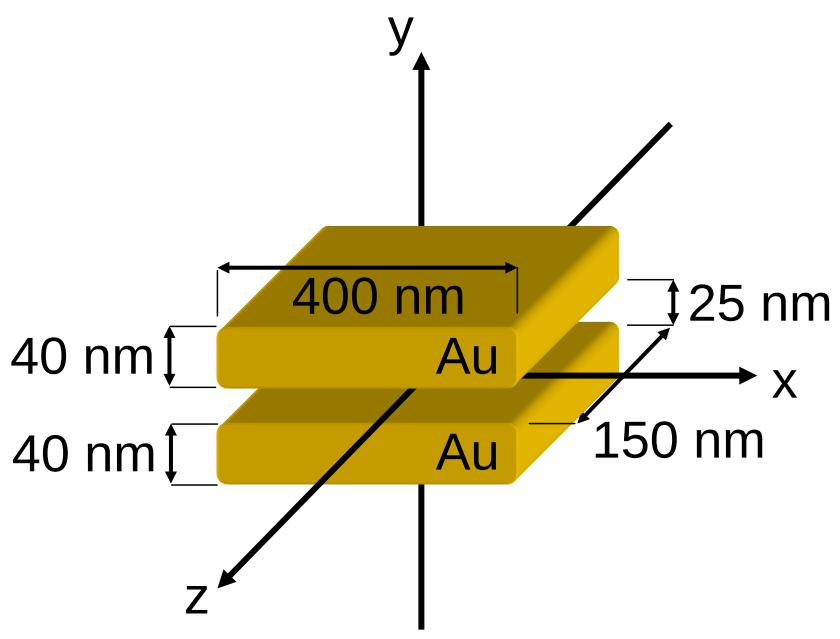

**Fig. 3: Sizes of the structure considered in the paper. In case of dipole-like nanoresonator, only one "brick" is left, while for the quadrupole-like nanoresonator both "bricks" are presented.**

In the case of the dipole-like nanoresonator, only one "brick" is considered, while for the quadrupole-like structure both "bricks" are presented. In this case one more value, namely optical coupling coefficient $\sigma$ (see [13] and (15)) can be straightforwardly evaluated by the measurements of the eigen frequency splitting (remind that the system shown in Fig. 3 exhibits resonance splitting due to the optical coupling between the "bricks").

The parameters of the QS are usually tabulated and can be easily found in the literature. The phase relaxation time $\tau_2$ is recalculated from the spontaneous emission spectrum bandwidth and for example for the quantum dots is about 100 fs. The energy relaxation time $\tau_1$ (which is usually measured in the tests [34]) is from several to several tens of nanosecond for different experimental realizations. To be precise, we assume hereafter $\tau_1 = 1ns$. The dipole moment of the quantum emitter (e.g. quantum dot) is also tabulated and can be found in the literature; hereafter we assume $\mu_{QD} = 2.5*10^{-17}(cgse)$, which in combination of the quantum dots concentration of approximately $10^{19}/cm^3$ provides gain of about $10^5/cm$.

It is convenient to compare the values of amplitude $A$ (actually, square of modulus $|A|^2$) in units of the saturation square amplitude $|A_{sat}|^2 = \frac{\hbar^2}{4\mu_{QD}^2\tau_1\tau_2}$, which corresponds to the saturation intensity $I_{sat} = \frac{c}{4\pi}|A|^2_{sat} \sim 1(KW/cm^2)$. This level of intensity is easily achievable with the lasers having low output power of about *1 mW* and appropriate focusing. If $A$ remains lower these values, the saturation effect can be neglected to the first approximation. It is methodologically correct to consider the influence of the external electric field in the regions of low (in compare with the saturation) powers $|A| << |A_{sat}|$, in the vicinity of the saturation power $|A| \sim |A_{sat}|$, and in the deep saturation region $|A| >> |A_{sat}|$. The peculiarities of the system dynamics (and the respective measurable data) allow concluding already on a qualitative level about values of the field acting on the quantum emitters.

The introduced in (14) the effective kinetic inductance of the nanoresonator $\chi$ can be evaluated based on the data about field enhancement by the nanoresonator. In fact, the variable $x$ has the same unit as the electric field $A$ ($\chi$ has units of frequency in square) and the relation $|x|N_{el}/|A|$ is the field enhancement factor, which can be easily calculated numerically for several frequencies around the resonance. After that $\chi$ can be fitted with the expression:

$$\chi = \left[|x|N_{el}/|A|\right](\omega) * \left(\left(\omega_0^2-\omega^2\right)^2+4\omega^2\gamma^2\right)^{1/2}/N_{el} \qquad (15)$$

Here $N_{el}$ is the effective number of electrons contributing to the field enhancement. We assume here $n_{freeel\,electrons}=5{,}9*10^{22}\left(cm^{-3}\right)$ and $V_{resonator}=400\times150\times40(nm^3)=2{,}4*10^{-15}(cm^3)$, which gives for the effective number of electrons $N_{el}=n_{freeel\,ectrons}*V_{resonator}=0{,}5*10^8$. In the developed here model $\chi$ is assumed to be constant, and thus far the deviation of (15) from the straight line estimates accuracy of the made approximation. For the case of resonance, and assuming reasonable field enhancement value of $FE=10^2$ we obtain:

$$\chi = FE*2\omega_0\gamma/N_{el} \approx 4\times10^{22}\left(s^{-2}\right) \qquad (16)$$

The same value can be estimated using the expressions for the electric field generated by a dipole in a near field zone. Starting from the harmonic oscillator equation for the electron displacement $\delta x$:

$$\frac{d^2(\delta x)}{dt^2}+2\gamma\frac{d(\delta x)}{dt}+\omega_0^2(\delta x)=\frac{q_{el}}{m_{el}}\mathrm{E} \qquad (17)$$

and introducing generated by the oscillating dipole field at the distance $R$ $x=\frac{q_{el}\delta x}{R^3}$, we get for the variable $x$ (see (14)):

$$\frac{d^2x}{dt^2}+2\gamma\frac{dx}{dt}+\omega_0^2x=\frac{q_{el}^2}{m_{el}R^3}\mathrm{E} \qquad (18)$$

which gives for the coefficient $\chi$ at $R=30\ nm$:

$$\chi = \frac{q_{el}^2}{m_{el} R^3} \sim 10^{25} \left( esu^2 / g / cm^3 \right) \tag{19}$$

here $q_{el} = -4{,}8*10^{-10}(esu)$ and $m_{el} = 9{,}1*10^{-28}(g)$ are the electron charge and mass, respectively. It is seen that (19) two orders of magnitude higher than evaluation (16). The reason for this deviation could be in overestimation of the number of electron really contributing to the plasmonic oscillations. Assuming that instead of all electrons in the nanoresonator only 1% ( $N_{el,eff} = 10^{-2} n_{freeelectrons} * V_{resonator} = 0{,}5*10^6$ ) participate in the oscillations we obtain order of magnitude matching between these two estimations. This important result will be used in other estimations; from the other side, it also indicates, that not all electrons are involved in the plasmonic dynamics.

The two coupling coefficients $\alpha_x$ and $\alpha_\rho$ can be evaluated only roughly. Up to now, we did not have a chance to compare our qualitative model with the rigorous full scale numerical calculations (it is nevertheless planned as a next step) and all provided below numbers have to be taken only as a rough approximation. Assuming that the field *E* in (14) is caused by the quantum dots placed at the distance *R* $E = \frac{N_{QD} \mu_{QD} \rho_{12}}{R^3}$ (here $N_{QD}$ is the number of the quantum dotsand $\mu_{QD} = 2.5*10^{-17}(cgse)$ ), the third equation in (11) becomes:

$$\begin{cases} \frac{d^2 x}{dt^2} + 2\gamma \frac{dx}{dt} + \omega_0^2 x = \frac{q_{el}^2}{m_{el} R^3} \frac{N_{QD} \mu_{QD} \rho_{12}}{R^3} = \alpha_\rho \rho_{12} \\ \alpha_\rho = \frac{q_{el}^2}{m_{el}} \frac{N_{QD} \mu_{QD}}{R^6} \end{cases} \tag{20}$$

If the quantum dots $QD_size = 3(nm) = 3*10^{-7}(cm)$ densely packed from one side of the nanoresonator at the effective distance *R*, one can roughly accept $N_{QD} = \frac{400 \times 150}{\pi (QD_size)^2} = 2000$.

$$\alpha_\rho = \frac{q_{el}^2}{m_{el}} \frac{N_{QD} \mu_{QD}}{R^6} \sim 2*10^{28} (statV / cm / s^2) \tag{21}$$

The meaning of the second coupling coefficient physically coincides with the dipole moment of the quantum emitter $\mu_{QD}$ (remind, that variable *x* is the field generated by one electron at the point of quantum emitter), but has to be multiplied by the number of electrons involved into the plasmonic oscillations:

$$\alpha_x = \mu_{QD} N_{el} \sim 10^{-11} (cgse) \qquad (22)$$

The choice of the particular numerical values for the numerical calculations was stipulated by the necessity to get an "observable effects". For example, for the spaser dynamics the "questionable" values are both coupling constants, while all other parameters could be pretty much independently determined. We have accepted for our calculations the values closed to (21) (22), and assumed maximum achievable population inversion $N_{0,\max} = 0{,}9$. Under these values and number of quantum dots $N_{QD} = 2000$ the threshold for the spaser generation and the stable spaser generation can be achieved. Anyway, comparison with the rigorous calculations and experimental data is mandatory to fix the model parameters and is a topic of another publication.

## 4. Extension on the case of metaatoms with magnetic response

Among possible applications of the nanolaser it was proposed to achieve generation using non emitting (dark) modes of the plasmonic resonators [30]. It was claimed that the lasing with the dark modes should have lower threshold, and consequently has to be achieved at lower pump levels [30]. Here a combination of cut wires (see Fig. 2(b) and Fig. 3) and interaction of this structure with the QS is considered. The system of two coupled oscillators possesses symmetric (dipole like) and asymmetric (quardupole-like) modes with different respective eigenfrequencies $\omega_{sym}$ and $\omega_{asym}$. The transition frequency of the QS $\omega_{21}$ can be adjusted in order to match the respective eigenfrequency and consequently provide maximum interaction efficiency. To be precise, we restrict our consideration by the object with the same geometric sizes as in [13] – see Fig. 3.

In order to elaborate the respective system of equations in analog with (14), it is necessary to substitute the single harmonic oscillator by two coupled harmonic oscillators, as it has been done for the double wires MAs in [13]. It is assumed also that only one nanoresonator is coupled with the QS, and the dynamics of the second nanoresonator is driven by the coupling with the first one; the slowly varying approximation remains the same (12), and the both coupled nanoresonators are assumed to be equivalent, i.e. eigenfrequencies $\omega_0$ and loss coefficients $\gamma$ for the both nanoresonators are the same. The resulting system of equations is:

$$\begin{cases} \dfrac{d\tilde{\rho}_{12}}{dt}+\tilde{\rho}_{12}\left(\dfrac{1}{\tau_2}+i\left(\omega-\omega_{21}\right)\right)=\dfrac{i\alpha_x \tilde{x}_1^{*}N}{\hbar}+\dfrac{i\mu_{QS}A^{*}N}{\hbar}+\xi_\rho \\ \dfrac{dN}{dt}+\dfrac{\left(N-N_0\right)}{\tau_1}=\dfrac{i\alpha_x\left(\tilde{x}_1\tilde{\rho}_{12}-\tilde{x}_1^{*}\tilde{\rho}_{12}^{*}\right)+i\mu_{QS}\left(A\tilde{\rho}_{12}-A^{*}\tilde{\rho}_{12}^{*}\right)}{2\hbar} \\ 2\left(\gamma-i\omega\right)\dfrac{d\tilde{x}_1}{dt}+\left(\omega_0^2-\omega^2-2i\omega\gamma\right)\tilde{x}_1+\sigma x_2=\alpha_\rho\tilde{\rho}_{12}^{*}+\chi A_1+\xi_{x1} \\ 2\left(\gamma-i\omega\right)\dfrac{d\tilde{x}_2}{dt}+\left(\omega_0^2-\omega^2-2i\omega\gamma\right)\tilde{x}_2+\sigma x_1=\chi A_2+\xi_{x2} \end{cases} \tag{23}$$

The phenomenological constant $\sigma$ describes coupling between the nanoresonators through the near field, and $A_{1,2}$ are the fields acting on the upper and lower nanoresonator. The symmetric and asymmetric oscillation modes (keeping in mind symmetric and asymmetric eigenfrequencies $\omega_{sym}=\sqrt{\omega_0^2+\sigma}$ and $\omega_{asym}=\sqrt{\omega_0^2-\sigma}$ respectively) can be straightforwardly introduced according to:

$$\begin{cases} m_s=\tilde{x}_1+\tilde{x}_2 \\ m_a=\tilde{x}_1-\tilde{x}_2 \end{cases} \tag{24}$$

In these new variables system (23) becomes:

$$\begin{cases} \dfrac{d\tilde{\rho}_{12}}{dt}+\tilde{\rho}_{12}\left(\dfrac{1}{\tau_2}+i\left(\omega-\omega_{21}\right)\right)=\dfrac{i\alpha_x\left(m^{*}_s+m^{*}_a\right)N}{2\hbar}+\dfrac{i\mu_{QS}A_1^{*}N}{\hbar}+\xi_\rho \\ \dfrac{dN}{dt}+\dfrac{\left(N-N_0\right)}{\tau_1}=\dfrac{i\alpha_x\left(\left(m_s+m_a\right)\tilde{\rho}_{12}-\left(m^{*}_s+m^{*}_a\right)\tilde{\rho}_{12}^{*}\right)+i\mu_{QS}\left(A_1\tilde{\rho}_{12}-A_1^{*}\tilde{\rho}_{12}^{*}\right)}{4\hbar} \\ 2\left(\gamma-i\omega\right)\dfrac{dm_s}{dt}+\left(\omega_0^2-\omega^2-2i\omega\gamma+\sigma\right)m_s=\alpha_\rho\tilde{\rho}_{12}^{*}+\chi\left(A_1+A_2\right)+\xi_{ms} \\ 2\left(\gamma-i\omega\right)\dfrac{dm_a}{dt}+\left(\omega_0^2-\omega^2-2i\omega\gamma-\sigma\right)m_a=\alpha_\rho\tilde{\rho}_{12}^{*}+\chi\left(A_1-A_2\right)+\xi_{ma} \end{cases} \tag{25}$$

In case of the absence of the external field $A$, system (25) describes dynamics of the “multipole spaser”, where one of the mode $m_s$ is coupled with the external space (bright mode) while the other one $m_a$ is to the first approximation not coupled with the far field zone (dark mode) and generates magnetic moment and magnetic field between the nanoresonators:

$$\begin{cases} \frac{d\tilde{\rho}_{12}}{dt}+\tilde{\rho}_{12}\left(\frac{1}{\tau_2}+i\left(\omega-\omega_{21}\right)\right)=\frac{i\alpha_x\left(m^*_s+m^*_a\right)N}{2\hbar}+\xi_\rho \\ \frac{dN}{dt}+\frac{\left(N-N_0\right)}{\tau_1}=\frac{i\alpha_x\left(\left(m_s+m_a\right)\tilde{\rho}_{12}-\left(m^*_s+m^*_a\right)\tilde{\rho}^*_{12}\right)}{4\hbar} \\ 2\left(\gamma_s-i\omega\right)\frac{dm_s}{dt}+\left(\omega_0^2-\omega^2-2i\omega\gamma_s+\sigma\right)m_s=\alpha_\rho\tilde{\rho}^*_{12}+\xi_{ms} \\ 2\left(\gamma_a-i\omega\right)\frac{dm_a}{dt}+\left(\omega_0^2-\omega^2-2i\omega\gamma_a-\sigma\right)m_a=\alpha_\rho\tilde{\rho}^*_{12}+\xi_{ma} \end{cases} \qquad (26)$$

The dark mode $m_a$ has less radiative losses $\gamma_a < \gamma_s$ and thus has lower generation threshold.

## 5. Elaboration of propagation equations

In this part a propagation equation for the EM wave in a MMs with MAs, consisting of coupled plasmonic resonators and QS will be elaborated. The approach is the natural extension of the multipole one, developed in [16] and allows us to investigate bulk properties of the MMs with gain and saturation type of nonlinearity and in the presence of a magnetic response.

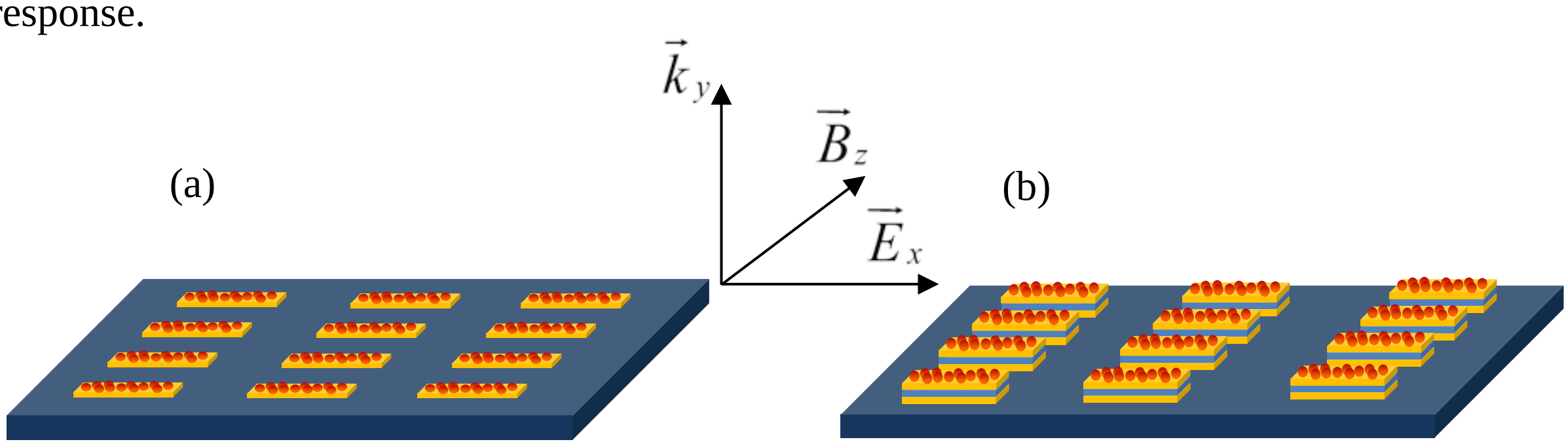

**Fig. 4: Schematic of the modelled active hybrid MM with QS: (a) – Dipole-like MA (one nanoresonator), (b) – Quadrupole-like MA (two coupled nano-resonators, separated by dielectric layer). The electric field is polarized along the long side of the wires. QSs are shown by the red circles on the top of the upper nanoresonators.**

This approach creates a solid basis which can be used for qualitative consideration of all most important problems appearing in case of consideration of the plane wave propagation in MMs with gain or, more generally, in case of MAs consisting of plasmonic nanoresonators, coupled with QS. The MMs with MAs depicted in Fig. 2 are considered. Elaboration of the propagation equation is the same as in case of the passive MAs [13]. The charge dynamics of the MAs becomes in this case more complicated and is described by system (14) for dipole-like MAs (Fig. 2(a)) or (25) for quadrupole-like MAs (Fig. 2(b)). From the other side, the propagation equation for the field and the calculation algorithm for the multipoles remain the same, and the full system of equations for the case of quadrupole-like particles (Fig. 2(b)) and CW wave is:

$$\begin{cases}\dfrac{\partial^2 E_x}{\partial y^2}+\dfrac{\omega^2}{c^2}\left(E_x+4\pi P_x(y,\rho_{12},\omega)\right)+\dfrac{i4\pi\omega}{c}\dfrac{\partial M_z(y,\rho_{12},\omega)}{\partial y}=0\\ \mathrm{P}_x(y,\rho_{12},\omega)=\eta q m_s-\dfrac{\partial Q_{xy}}{\partial y}\\ Q_{xy}(y,\rho_{12},\omega)=\eta q y_1 m_a\\ \mathrm{M}_z(y,\rho_{12},\omega)=\dfrac{i\omega\eta q y_1}{c}m_a\\ \dfrac{d\tilde{\rho}_{12}}{dt}+\tilde{\rho}_{12}\left(\dfrac{1}{\tau_2}+i\left(\omega-\omega_{21}\right)\right)=\dfrac{i\alpha_x\left(m^*_s+m^*_a\right)N}{2\hbar}+\dfrac{i\mu_{QS}E^*_{x,1}N}{\hbar}+\xi_\rho\\ \dfrac{dN}{dt}+\dfrac{\left(N-N_0\right)}{\tau_1}=\dfrac{i\alpha_x\left(\left(m_s+m_a\right)\tilde{\rho}_{12}-\left(m^*_s+m^*_a\right)\tilde{\rho}^*_{12}\right)+2i\mu_{QS}\left(E_{x,1}\tilde{\rho}_{12}-E^*_{x,1}\tilde{\rho}^*_{12}\right)}{4\hbar}\\ 2\left(\gamma-i\omega\right)\dfrac{dm_s}{dt}+\left(\omega_0^2-\omega^2-2i\omega\gamma+\sigma\right)m_s=\alpha_\rho\tilde{\rho}^*_{12}+\chi\left(E_{x,1}+E_{x,2}\right)+\xi_{ms}\\ 2\left(\gamma-i\omega\right)\dfrac{dm_a}{dt}+\left(\omega_0^2-\omega^2-2i\omega\gamma-\sigma\right)m_a=\alpha_\rho\tilde{\rho}^*_{12}+\chi\left(E_{x,1}-E_{x,2}\right)+\xi_{ma}\end{cases}\tag{27}$$

here $A$ is the electric field in the propagating wave, $A_1$ and $A_2$ are the electric field at upper and lower nanowires respectively, $P_x$ is the medium polarizability caused by a dipole and quadrupole contributions, while $Q$ and $M$ describe impact of higher order multipoles, giving rise the MMs effects. In this equation all multipole terms are functions of the non-diagonal element $\rho_{12}$, which comes to the multipole moments through the interaction term in the mode dynamics equations of system (19). For the case of dipole-like MAs (Fig. 2(a)) system (27) is reduced to:

$$\begin{cases}\dfrac{\partial^2 E_x}{\partial y^2}+\dfrac{\omega^2}{c^2}\left(E_x+4\pi P_x(y,\rho_{12},\omega)\right)=0\\ \mathrm{P}_x(y,\rho_{12},\omega)=\eta q m_s\\ 2\left(\gamma-i\omega\right)\dfrac{dm_s}{dt}+\left(\omega_0^2-\omega^2-2i\omega\gamma\right)m_s=\alpha_\rho\tilde{\rho}^*_{12}+2\chi E_x+\xi_{ms}\\ \dfrac{d\tilde{\rho}_{12}}{dt}+\tilde{\rho}_{12}\left(\dfrac{1}{\tau_2}+i\left(\omega-\omega_{21}\right)\right)=\dfrac{i\alpha_x m^*_s N}{2\hbar}+\xi_\rho\,\dfrac{i\mu_{QS}E^*_x N}{\hbar}\\ \dfrac{dN}{dt}+\dfrac{\left(N-N_0\right)}{\tau_1}=\dfrac{i\alpha_x\left(m_s\tilde{\rho}_{12}-m^*_s\tilde{\rho}^*_{12}\right)+2i\mu_{QS}\left(E_x\tilde{\rho}_{12}-E^*_x\tilde{\rho}^*_{12}\right)}{4\hbar}\end{cases}\tag{28}$$

Obviously the process of the loss compensation depends significantly on the degree of coupling of the emitters and MAs – the stronger coupling is, the more significantly charge dynamics in MA is changed, and in turn the more significantly effective optical properties will be changed. It is also clear, that in a real experimental realization the MM will contain both types of emitters, coupled and uncoupled, which has to be incorporated into the model.

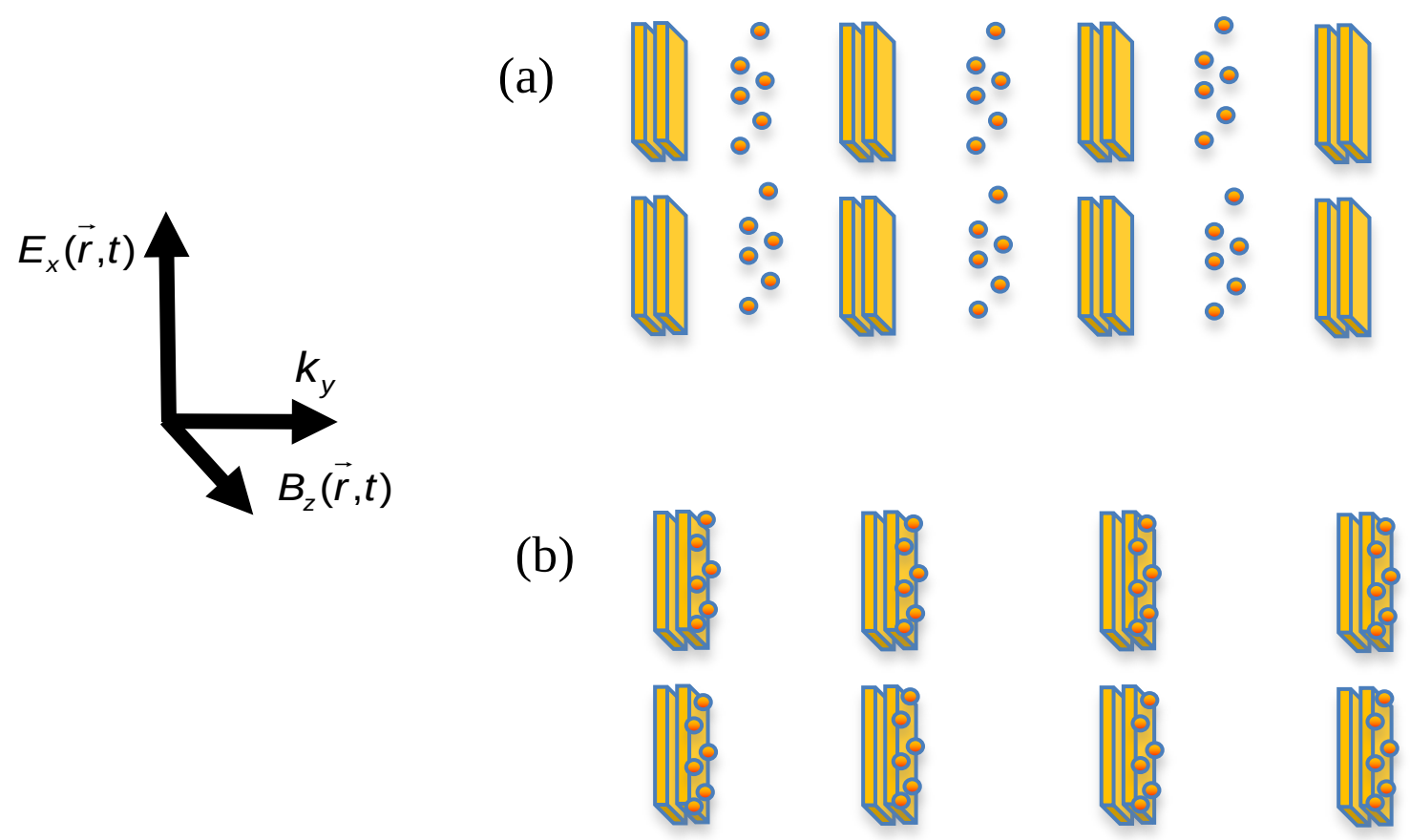

**Fig. 5: Schematic representation of the plane wave propagation in the metamaterial with (a) uncoupled and (b) coupled QSs.**

Here, we assume that total averaged concentration of the emitters (number of the emitters in an elementary volume, containing one MA) is kept constant for all cases considered below. In contrast, the particular emitters coupled to the MAs $\delta$ can vary from zero ($\delta = 0$, no coupled emitters) to one ($\delta = 1$, all emitters are coupled); both of them contribute to the polarizability:

$$
\begin{cases}
\dfrac{\partial^2 E_x}{\partial y^2}+\dfrac{\omega^2}{c^2}\left(E_x+4\pi P_x(y,\rho_{12},\omega)\right)+\dfrac{i4\pi\omega}{c}\dfrac{\partial M_z(y,\rho_{12},\omega)}{\partial y}=0 \\
\mathrm{P}_x(y,\rho_{12},\omega)=\eta q m_s-\dfrac{\partial Q_{xy}}{\partial y}+\eta_{QS}\left(1-\delta\right)\mu_{QS}\tilde{\rho}^*_{12,un}+\eta_{QS}\delta\mu_{QS}\tilde{\rho}^*_{12,c} \\
Q_{xy}(y,\rho_{12},\omega)=\eta q y_1 m_a \\
\mathrm{M}_z(y,\rho_{12},\omega)=\dfrac{i\omega\eta q y_1}{c}m_a \\
\\
\dfrac{d\tilde{\rho}_{12,un}}{dt}+\tilde{\rho}_{12,un}\left(\dfrac{1}{\tau_2}+i\left(\omega-\omega_{21}\right)\right)=\dfrac{i\mu_{QS}E^*_{x,1}N_{un}}{\hbar}+\xi_\rho \\
\dfrac{dN_{un}}{dt}+\dfrac{\left(N_{un}-N_0\right)}{\tau_1}=\dfrac{i\mu_{QS}\left(E_{x,1}\tilde{\rho}_{12,un}-E^*_{x,1}\tilde{\rho}^*_{12,un}\right)}{2\hbar} \\
\\
\dfrac{d\tilde{\rho}_{12,c}}{dt}+\tilde{\rho}_{12,c}\left(\dfrac{1}{\tau_2}+i\left(\omega-\omega_{21}\right)\right)=\dfrac{i\alpha_x\left(m^*_s+m^*_a\right)N}{2\hbar}+\dfrac{i\mu_{QS}E^*_{x,1}N_c}{\hbar}+\xi_\rho \\
\dfrac{dN_c}{dt}+\dfrac{\left(N_c-N_0\right)}{\tau_1}=\dfrac{i\alpha_x\left(\left(m_s+m_a\right)\tilde{\rho}_{12,c}-\left(m^*_s+m^*_a\right)\tilde{\rho}^*_{12,c}\right)+2i\mu_{QS}\left(E_{x,1}\tilde{\rho}_{12,c}-E^*_{x,1}\tilde{\rho}^*_{12,c}\right)}{4\hbar} \\
\\
2\left(\gamma-i\omega\right)\dfrac{dm_s}{dt}+\left(\omega_0^2-\omega^2-2i\omega\gamma+\sigma\right)m_s=\delta\alpha_\rho\tilde{\rho}^*_{12,c}+\chi\left(E_{x,1}+E_{x,2}\right)+\xi_{ms} \\
2\left(\gamma-i\omega\right)\dfrac{dm_a}{dt}+\left(\omega_0^2-\omega^2-2i\omega\gamma-\sigma\right)m_a=\delta\alpha_\rho\tilde{\rho}^*_{12,c}+\chi\left(E_{x,1}-E_{x,2}\right)+\xi_{ma}
\end{cases}
\tag{29}
$$

Here $\tilde{\rho}_{12,un}$ and $N_{un}$ describe dynamics of the uncoupled emitters, and $\tilde{\rho}_{12,un}$ and $N_{un}$ are responsible for the coupled emitter dynamics respectively. For the case of simplest dipole-like MA system (29) becomes:

$$
\begin{cases}
\dfrac{\partial^2 E_x}{\partial y^2}+\dfrac{\omega^2}{c^2}\left(E_x+4\pi P_x(y,\rho_{12},\omega)\right)=0 \\
\mathrm{P}_x(y,\rho_{12},\omega)=\eta q m_s+\eta_{QS}\left(1-\delta\right)\mu_{QS}\tilde{\rho}^*_{12,un}+\eta_{QS}\delta\mu_{QS}\tilde{\rho}^*_{12,c} \\
\\
\dfrac{d\tilde{\rho}_{12,un}}{dt}+\tilde{\rho}_{12,un}\left(\dfrac{1}{\tau_2}+i\left(\omega-\omega_{21}\right)\right)=\dfrac{i\mu_{QS}E^*_x N_{un}}{\hbar}+\xi_\rho \\
\dfrac{dN_{un}}{dt}+\dfrac{\left(N_{un}-N_0\right)}{\tau_1}=\dfrac{i\mu_{QS}\left(E_x\tilde{\rho}_{12,un}-E^*_x\tilde{\rho}^*_{12,un}\right)}{2\hbar} \\
\dfrac{d\tilde{\rho}_{12,c}}{dt}+\tilde{\rho}_{12,c}\left(\dfrac{1}{\tau_2}+i\left(\omega-\omega_{21}\right)\right)=\dfrac{i\alpha_x m^*_s N}{2\hbar}+\dfrac{i\mu_{QS}E^*_x N_c}{\hbar}+\xi_\rho \\
\dfrac{dN_c}{dt}+\dfrac{\left(N_c-N_0\right)}{\tau_1}=\dfrac{i\alpha_x\left(m_s\tilde{\rho}_{12,c}-m^*_s\tilde{\rho}^*_{12,c}\right)+2i\mu_{QS}\left(E_x\tilde{\rho}_{12,c}-E^*_x\tilde{\rho}^*_{12,c}\right)}{4\hbar} \\
2\left(\gamma-i\omega\right)\dfrac{dm_s}{dt}+\left(\omega_0^2-\omega^2-2i\omega\gamma+\sigma\right)m_s=\delta\alpha_\rho\tilde{\rho}^*_{12,c}+2\chi E_x+\xi_{ms}
\end{cases}
\tag{30}
$$

When all emitters are placed between MAs and are free of coupling ($\delta=0$, see Fig. 4(a)), the problem is reduced to the propagation of a plane wave with loss and gain, where gain dynamics and loss (plasmonic) dynamics are separated from each other:

$$
\begin{cases}
\dfrac{\partial^2 E_x}{\partial y^2}+\dfrac{\omega^2}{c^2}\left(E_x+4\pi P_x(y,\rho_{12},\omega)\right)+\dfrac{i4\pi\omega}{c}\dfrac{\partial M_z(y,\rho_{12},\omega)}{\partial y}=0 \\
\mathrm{P}_x(y,\rho_{12},\omega)=\eta q m_s-\dfrac{\partial Q_{xy}}{\partial y}+\eta_{QS}\mu_{QS}\tilde{\rho}^*_{12,un} \\
Q_{xy}(y,\rho_{12},\omega)=\eta q y_1 m_a \\
\mathrm{M}_z(y,\rho_{12},\omega)=\dfrac{i\omega\eta q y_1}{c}m_a \\
\\
\dfrac{d\tilde{\rho}_{12,un}}{dt}+\tilde{\rho}_{12,1}\left(\dfrac{1}{\tau_2}+i\left(\omega-\omega_{21}\right)\right)=\dfrac{i\mu_{QS}E^*_{x,1}N_{un}}{\hbar}+\xi_\rho \\
\dfrac{dN_{un}}{dt}+\dfrac{\left(N_{un}-N_0\right)}{\tau_1}=\dfrac{i\mu_{QS}\left(E_{x,1}\tilde{\rho}_{12,un}-E^*_{x,1}\tilde{\rho}^*_{12,un}\right)}{2\hbar} \\
\\
2\left(\gamma-i\omega\right)\dfrac{dm_s}{dt}+\left(\omega_0^2-\omega^2-2i\omega\gamma+\sigma\right)m_s=\chi\left(E_{x,1}+E_{x,2}\right)+\xi_{ms} \\
2\left(\gamma-i\omega\right)\dfrac{dm_a}{dt}+\left(\omega_0^2-\omega^2-2i\omega\gamma-\sigma\right)m_a=\chi\left(E_{x,1}-E_{x,2}\right)+\xi_{ma}
\end{cases}
\tag{31}
$$

and no new physical phenomena are expected.

## 6. Propagation of a plane wave in a metamaterial with dipole-like metaatoms

First, the case of dipole-like MAs (30) is considered in the case of CW operation without noise terms, namely:

$$\begin{cases}
\dfrac{\partial^2 E_x}{\partial y^2}+\dfrac{\omega^2}{c^2}\left(E_x+4\pi P_x(y,\rho_{12},\omega)\right)=0 \\
\mathrm{P}_x(y,\rho_{12},\omega)=\eta q m_s+\eta_{QS}\left(1-\delta\right)\mu_{QS}\tilde{\rho}^*_{12,un}+\eta_{QS}\delta\mu_{QS}\tilde{\rho}^*_{12,c} \\
\tilde{\rho}_{12,un}=\dfrac{i\mu_{QS}E^*_x N_{un}}{2\hbar R_\rho} \\
N_{un}-N_0=\dfrac{i\tau_1\mu_{QS}\left(E_x\tilde{\rho}_{12,un}-E^*_x\tilde{\rho}^*_{12,un}\right)}{4\hbar} \\
\tilde{\rho}_{12,c}=\dfrac{iN_c}{2\hbar R_\rho}\left(\alpha_x m^*{}_s+2\mu_{QS}E^*_x\right) \\
N_c-N_0=\dfrac{i\tau_1}{4\hbar}\left(\alpha_x\left(m{}_s\tilde{\rho}_{12,c}-m^*{}_s\tilde{\rho}^*_{12,c}\right)+2\mu_{QS}\left(E_x\tilde{\rho}_{12,c}-E^*_x\tilde{\rho}^*_{12,c}\right)\right) \\
m_s=\dfrac{\delta\alpha_\rho\tilde{\rho}^*_{12,c}+2\chi E_x}{R_s} \\
R_\rho=\left(1+i\left(\omega-\omega_{21}\right)\tau_2\right) \\
R_s=\left(\omega_0^2-\omega^2-2i\omega\gamma+\sigma\right)
\end{cases} \tag{32}$$

The Helmholz propagation equation for the electric field can be transformed using Slowly Varying Approximation (SVA):

$$\begin{aligned}
&E_x\left(y\right)=A_x\left(y\right)\exp\left(iky\right)+\frac{\omega^2}{c^2}\left(E_x(y,\omega)+4\pi P_x(y,\omega)\right)=0 \\
&\Rightarrow\frac{\partial^2 E_x(y,\omega)}{\partial y^2}=\frac{\partial^2 A_x(y,\omega)}{\partial y^2}+2ik\frac{\partial A_x(y,\omega)}{\partial y}-k^2A_x\approx 2ik\frac{\partial A_x(y,\omega)}{\partial y}-k^2A_x
\end{aligned} \tag{33}$$

Note, that all equations for $\tilde{\rho}_{12,un}$ , $\tilde{\rho}^*_{12,c}$ , $N_{un}$ , $N_c$ , and $m_s$ remain invariant under transformation:

$$\begin{cases} E_x \to A_x \exp(iky) \\ \tilde{\rho}_{12,un} \to \tilde{\rho}_{12,un} \exp(-iky) \\ \tilde{\rho}_{12,c} \to \tilde{\rho}_{12,c} \exp(-iky) \\ m_s \to m_s \exp(iky) \end{cases} \tag{34}$$

Combination of (33) and (34) finally results in:

$$\begin{cases} ik\dfrac{\partial A_x(y,\omega)}{\partial y} + 2\pi\dfrac{\omega^2}{c^2}\left( \delta\left( q\eta\dfrac{\alpha_\rho}{R_s} + \eta_\rho\mu_{QS} \right)\rho^*_{12,c}(\omega, A_x) + \eta_\rho\mu_{QS}(1-\delta)\rho^*_{12,un}(\omega, A_x) \right) = 0 \\ \tilde{\rho}_{12,un} = \dfrac{i\mu_{QS}A^*_x N_{un}}{2\hbar R_\rho} \\ N_{un} - N_0 = \dfrac{i\tau_1\mu_{QS}\left( A_x\tilde{\rho}_{12,un} - A^*_x\tilde{\rho}^*_{12,un} \right)}{4\hbar} \\ \tilde{\rho}_{12,c} = \dfrac{iN_c}{2\hbar R_\rho}\left( \alpha_x m^*_s + 2\mu_{QS}A^*_x \right) \\ N_c - N_0 = \dfrac{i\tau_1}{4\hbar}\left( \alpha_x\left( m_s\tilde{\rho}_{12,c} - m^*_s\tilde{\rho}^*_{12,c} \right) + 2\mu_{QS}\left( A_x\tilde{\rho}_{12,c} - A^*_x\tilde{\rho}^*_{12,c} \right) \right) \\ m_s = \dfrac{\delta\alpha_\rho\tilde{\rho}^*_{12,c} + 2\chi A_x}{R_x} \end{cases} \tag{35}$$

Where the following substitution has been made $k^2 = \dfrac{\omega^2}{c^2}\left( 1 + 4\pi q\eta\dfrac{\chi}{R_s} \right)$. System (35) is the master system of equations to describe propagation of the plane wave in a MM with MAs, consisting of plasmonic nanoresonators coupled with QS and free QS embedded in the matrix of the MM itself. In the case of Quantum Dots, for example, as QS, (35) describes loss compensation and respective optical properties of the MM.

Solutions of (35), especially transition region, can only be found numerically. Nevertheless, even before the solutions are demonstrated, one can qualitatively predict different operation modes, where stable and unstable operation modes of the spaser have been analyzed [15].

**Loss compensation by completely uncoupled QS ($\delta = 0$)**

In this case (35) becomes:

$$\begin{cases} ik\dfrac{\partial A_x(y,\omega)}{\partial y}+2\pi\dfrac{\omega^2}{c^2}\eta_\rho\mu_{QS}\rho^*_{12,un}(\omega,A_x)=0 \\ \tilde{\rho}_{12,un}=\dfrac{i\mu_{QS}A_x^*N_{un}}{2\hbar R_\rho} \\ N_{un}-N_0=\dfrac{i\tau_1\mu_{QS}\left(A_x\tilde{\rho}_{12,un}-A_x^*\tilde{\rho}^*_{12,un}\right)}{4\hbar} \\ m_s=\dfrac{2\chi A_x}{R_x} \end{cases} \tag{36}$$

Solution of (36) describes rather trivial loss compensation in a media with dipoles and will not be considered here; more information about this kind of problems can be found in any textbook about optical amplifiers, see for example [31]. Under appropriate choice of parameters (QS concentration and pump) plane wave stabilizes own intensity by saturation and propagates stably.

The results of the solution of the propagation equations (36) are presented in Fig. 6, where amplitude $A_x$ of the propagating wave is presented in color grade as a function of frequency and propagation length.

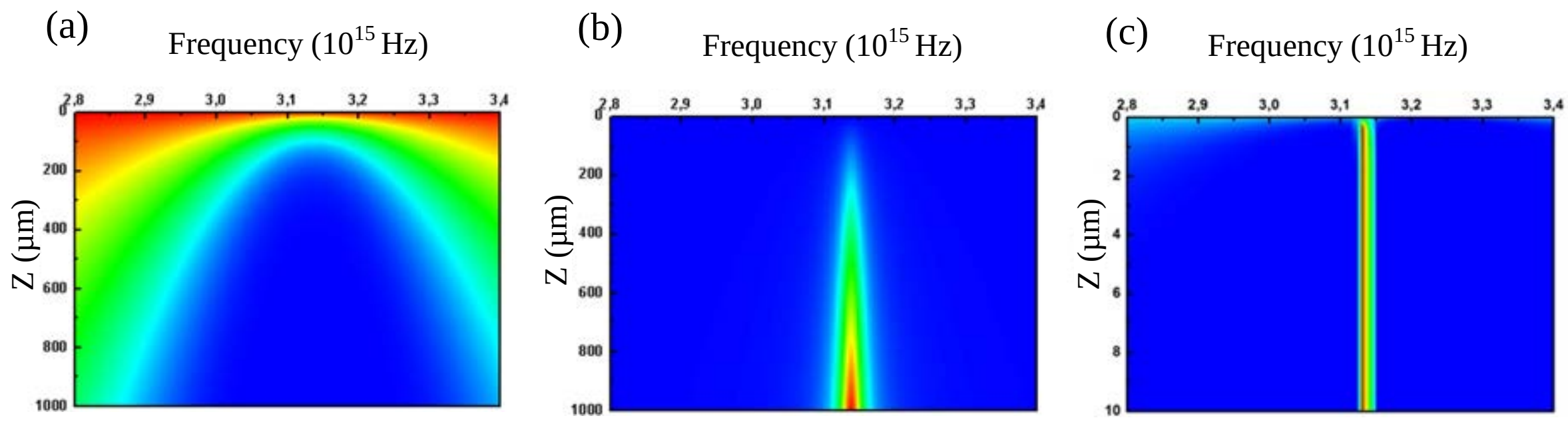

**Fig. 6: (a) Propagation of the plane wave in the case of MM with dipole-like MAs without QS, blue color corresponds to lower amplitudes. Peak of the losses corresponds to the resonance frequency for a single nanoresonator $\omega_0 = 3{,}14*10^{15}Hz$. (b) Propagation of the plane wave in media filled by pumped QS with the resonant frequency $\omega_{21} = 3{,}14*10^{15}Hz$. (c) Propagation of the plane wave in the case of completely uncoupled QS $\delta = 0$ in MM with dipole-like MAs, $\omega_0 = \omega_{21} = 3{,}14*10^{15}Hz$. Parameters (QS concentration and pump) are chosen in order to provide full loss compensation, which is achieved around QS peak gain frequency.**

The parameters of the MM (nanoresonator concentration, QS concentration, pump etc.) have been chosen in order to reach full compensation, so that amount of gain provided to the plane wave is enough to compensate the losses. The stable propagation can be achieved around the central gain frequency of the QS, which is chosen the same as the resonant frequency of the nanoresonators.

**Loss compensation by completely coupled QS ($\delta = 1$)**

In this case (35) is reduced to the system, which describes the situation considered, for example, in [12]:

$$\begin{cases} ik\dfrac{\partial A_x(y,\omega)}{\partial y} + 2\pi\dfrac{\omega^2}{c^2}\left(q\eta\dfrac{2\alpha_\rho}{R_x} + \eta_\rho\mu_{QS}\right)\rho^*_{12,c}(\omega, A_x) = 0 \\ \tilde{\rho}_{12,c} = \dfrac{iN_c}{2\hbar R_\rho}\left(\alpha_x m^*{}_s + 2\mu_{QS}A_x^*\right) \\ N_c - N_0 = \dfrac{i\tau_1}{4\hbar}\left(\alpha_x\left(m{}_s\tilde{\rho}_{12,c} - m^*{}_s\tilde{\rho}^*_{12,c}\right) + 2\mu_{QS}\left(A_x\tilde{\rho}_{12,c} - A_x^*\tilde{\rho}^*_{12,c}\right)\right) \\ m_s = \dfrac{\delta\alpha_\rho\tilde{\rho}^*_{12,c} + 2\chi A_x}{R_x} \end{cases} \tag{37}$$

In this case full loss compensation coincides with the threshold of the spaser generation – which means, that the spaser’s own dynamics will be affected by the external wave, resulting in potentially unstable operation [15]. To this extent, the results about instability of operation of the MM under full loss compensation correspond to the conclusion of [12]. Nevertheless, stable operation is still possible for not completely coupled case. On the other hand, it is hard to believe that under real experimental situations all emitters appear to be coupled with the plasmonic nanoresonators. Therefore, an experimental realization of the case $\delta = 1$ and respective instability prognoses remain rather exotic ones.

The solutions of propagation equations (37) are presented in Fig. 7. In this case peak luminescence coincided (as before) with the nanoresonator frequency $\omega_{21} = \omega_0$ and the complex structure of the spectrum is caused by strong coupling between the QS and the nanoresonators. It should be emphasized that for the same parameters, but without coupling the amplitude of the propagating wave cannot be stabilized for any frequency. It proves in turn that the coupling between the QS and the nanoresonator reduces significantly an amount of QS necessary for the full loss compensation and is crucial for the experimentally achievable full loss compensation in MMs.

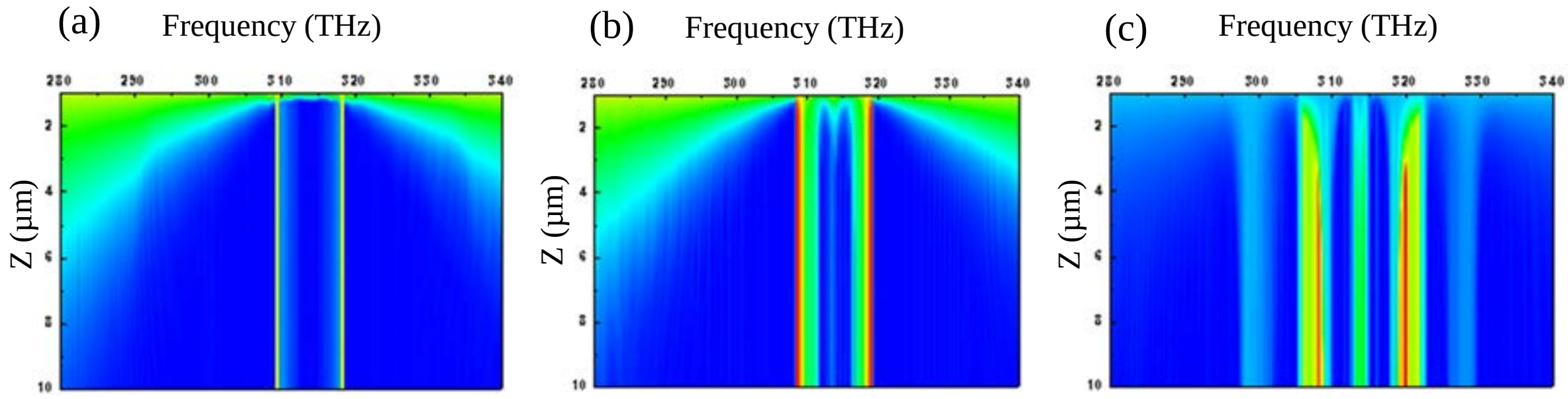

**Fig. 7: Propagation of the plane wave in the case of totally coupled QDs in MM with dipole-like MAs. The amount of QD is fixed, $\omega_{12} = \omega_0 = 314THz$, $N_0$ (the pump rate) is varied 0.93 0.97 0.98 for (a), (b), and (c) respectively. One can see that there are certain frequencies where losses are totally compensated and stable propagation is possible. Amount of such frequencies increases as the pump rate increases**.

### Loss compensation by partially coupled QS ($0 < \delta < 1$)

It has been shown that the coupling of plasmonic nanoresonators to QSs increases amplification and offers a preferable realization of loss compensation [32]. As just discussed, a fully coupled realization tends to result in unstable operation [15]. In the intermediate case (35) the number of coupled emitters is almost sufficient to cross the generation threshold. The rest of losses will be compensated by free emitters. This seems to be perfect solution and sets a clear strategy for the design of the MMs with fully compensated losses. The solution, qualitatively looks similar to the one presented in Fig. 7 and is not shown here. The main difference is in the wider region of parameters for stable propagation (the parameters correspond to the stability region). The full analysis of stability of the plane wave propagation in MMs with gain is a subject for further investigation.

## 7. Propagation of plane wave in metamaterial with quadrupole-like metaatoms

In summary, consideration of dipole-like MAs and the respective MMs reveal mainly two new physical phenomena, namely:

- Enhancement of the amplification processes due to the appropriate positioning of the emitter in the vicinity of the plasmonic nanoresonators and coupling between the emitters and nanoresonators;
- Possibly unstable plane wave propagation in MMs where all emitters are coupled with the plasmonic nanoresonators, which requires non zero free (uncoupled) emitters which contribute to the loss compensation.

The most intriguing property of MMs, namely magnetic response in the optical frequency domain, cannot be considered in the framework of the dipole-like MAs and requires a full consideration of multipoles, namely magnetic dipole and quadrupole terms. In such a case, system (29) will be reduced to the system of equation for SVA in analogy with the consideration of the dipole-like MAs. Firstly, system (29) for CW operation and without stochastic terms becomes:

$$
\begin{cases}
\frac{\partial^2 E_x}{\partial y^2}+\frac{\omega^2}{c^2}\left(E_x+4\pi P_x(y,\rho_{12},\omega)\right)+\frac{i4\pi\omega}{c}\frac{\partial M_z(y,\rho_{12},\omega)}{\partial y}=0 \\
\mathrm{P}_x(y,\rho_{12},\omega)=\eta q m_s-\frac{\partial Q_{xy}}{\partial y}+\eta_{QS}(1-\delta)\mu_{QS}\tilde{\rho}^*_{12,un}+\eta_{QS}\delta\mu_{QS}\tilde{\rho}^*_{12,c} \\
Q_{xy}(y,\rho_{12},\omega)=\eta q y_1 m_a \\
\mathrm{M}_z(y,\rho_{12},\omega)=\frac{i\omega\eta q y_1}{c}m_a \\
\tilde{\rho}_{12,un}=\frac{i\mu_{QS}\tau_2 E^*_{x,1}N_{un}}{\hbar R_\rho} \\
N_{un}-N_0=\frac{i\mu_{QS}\tau_1\left(E_{x,1}\tilde{\rho}_{12,un}-E^*_{x,1}\tilde{\rho}^*_{12,un}\right)}{2\hbar} \\
\tilde{\rho}_{12,c}=\frac{i\tau_2}{2\hbar R_\rho}\left(\alpha_x\left(m^*_s+m^*_a\right)N+2\mu_{QS}E^*_{x,1}N_c\right) \\
N_c-N_0=\frac{i\tau_1}{4\hbar}\left(\alpha_x\left(\left(m_s+m_a\right)\tilde{\rho}_{12,c}-\left(m^*_s+m^*_a\right)\tilde{\rho}^*_{12,c}\right)+2i\mu_{QS}\left(E_{x,1}\tilde{\rho}_{12,c}-E^*_{x,1}\tilde{\rho}^*_{12,c}\right)\right) \\
m_s=\frac{\delta\alpha_\rho\tilde{\rho}^*_{12,c}+\chi\left(E_{x,1}+E_{x,2}\right)}{R_s} \\
m_a=\frac{\delta\alpha_\rho\tilde{\rho}^*_{12,c}+\chi\left(E_{x,1}-E_{x,2}\right)}{R_a} \\
R_\rho=\left(1+i\left(\omega-\omega_{21}\right)\tau_2\right) \\
R_s=\left(\omega_0^2-\omega^2-2i\omega\gamma+\sigma\right) \\
R_a=\left(\omega_0^2-\omega^2-2i\omega\gamma-\sigma\right)
\end{cases}
\qquad (38)
$$

In contrast to the previous case, with just dipole-like MAs, one is required to take into account both quadrupole and magnetic dipole moments in the propagation equation, which are included in the equation through the first derivatives. Moreover, $E_{x,1}$ and $E_{x,2}$ are the fields driving upper and lower nanowires respectively. Following [13] it is assumed that:

$$E_{x,1}=E_x \qquad (39)$$

and:

$$E_{x,2}=E_x\exp\left(iky_1\right) \qquad (40)$$

$k$ is the wave vector, and $y_1$ is the distance between nanowires (see [13]). As in [13], the wave vector must be found by consideration of the propagation equation. Substituting (39) and (40) into (38), and taking into account that $ky_1 << 1$, system (38) becomes:

$$\begin{cases}
\frac{\partial^2 E_x}{\partial y^2}+\frac{\omega^2}{c^2}\left(E_x+4\pi P_x(y,\rho_{12},\omega)\right)+\frac{i4\pi\omega}{c}\frac{\partial M_z(y,\rho_{12},\omega)}{\partial y}=0 \\
P_x(y,\rho_{12},\omega)=\eta q m_s-\frac{\partial Q_{xy}}{\partial y}+\eta_{QS}(1-\delta)\mu_{QS}\tilde{\rho}^*_{12,un}+\eta_{QS}\delta\mu_{QS}\tilde{\rho}^*_{12,c} \\
Q_{xy}(y,\rho_{12},\omega)=\eta q y_1 m_a \\
M_z(y,\rho_{12},\omega)=\frac{i\omega\eta q y_1}{c}m_a \\
\tilde{\rho}_{12,un}=\frac{i\mu_{QS}\tau_2 E^*_x N_{un}}{\hbar R_\rho} \\
N_{un}-N_0=\frac{i\mu_{QS}\tau_1\left(E_x\tilde{\rho}_{12,un}-E^*_x\tilde{\rho}^*_{12,un}\right)}{2\hbar} \\
\tilde{\rho}_{12,c}=\frac{i\tau_2}{2\hbar R_\rho}\left(\alpha_x\left(m^*_s+m^*_a\right)N+2\mu_{QS}E^*_x N_c\right) \\
N_c-N_0=\frac{i\tau_1}{4\hbar}\left(\alpha_x\left(\left(m_s+m_a\right)\tilde{\rho}_{12,c}-\left(m^*_s+m^*_a\right)\tilde{\rho}^*_{12,c}\right)+2i\mu_{QS}\left(E_x\tilde{\rho}_{12,c}-E^*_x\tilde{\rho}^*_{12,c}\right)\right) \\
m_s=\frac{\delta\alpha_\rho\tilde{\rho}^*_{12,c}+2\chi E_x\left(1-\left(ky_1\right)^2\right)}{R_s} \\
m_a=\frac{\delta\alpha_\rho\tilde{\rho}^*_{12,c}+2i\chi E_x\left(ky_1\right)}{R_a}
\end{cases} \tag{41}$$

Reduction of (41) to the SVA can be done following the same steps as (33), (34) with the additional assumption that $\frac{\partial}{\partial y}Q_{xy}(y,\rho_{12},\omega)\rightarrow ik\,\eta q y_1 m_a$ and $M_z(y,\rho_{12},\omega)\rightarrow ik\frac{i\omega\eta q y_1}{c}m_a$ ; wave vector is also set to $k^2=\frac{\omega^2}{c^2}\left(1+4\pi q\eta\frac{\chi}{R_s}\right)$. Finally (41) becomes:

$$
\begin{cases}
ik\dfrac{\partial A_x(y,\omega)}{\partial y}+2\pi\dfrac{\omega^2}{c^2}\delta\left[\left(q\eta\alpha_\rho\left(\dfrac{1}{R_s}-\dfrac{2iky_1}{R_a}\right)+\eta_\rho\mu_{QS}\right)\rho^*_{12,c}(\omega,A_x)+\right.\\
\qquad\left.+\eta_\rho\mu_{QS}(1-\delta)\rho^*_{12,un}(\omega,A_x)+2\chi E_x(ky_1)^2\left(\dfrac{2}{R_a}-\dfrac{1}{R_s}\right)\right]=0\\
\tilde{\rho}_{12,un}=\dfrac{i\mu_{QS}\tau_2E_x^*N_{un}}{\hbar R_\rho}\\
N_{un}-N_0=\dfrac{i\mu_{QS}\tau_1\left(E_x\tilde{\rho}_{12,un}-E_x^*\tilde{\rho}^*_{12,un}\right)}{2\hbar}\\
\tilde{\rho}_{12,c}=\dfrac{i\tau_2}{2\hbar R_\rho}\left(\alpha_x\left(m^*_s+m^*_a\right)N+2\mu_{QS}E_x^*N_c\right)\\
N_c-N_0=\dfrac{i\tau_1}{4\hbar}\left(\alpha_x\left((m_s+m_a)\tilde{\rho}_{12,c}-(m^*_s+m^*_a)\tilde{\rho}^*_{12,c}\right)+2i\mu_{QS}\left(E_x\tilde{\rho}_{12,c}-E_x^*\tilde{\rho}^*_{12,c}\right)\right)\\
m_s=\dfrac{\delta\alpha_\rho\tilde{\rho}^*_{12,c}+2\chi E_x\left(1-(ky_1)^2\right)}{R_s}\\
m_a=\dfrac{\delta\alpha_\rho\tilde{\rho}^*_{12,c}+2i\chi E_x(ky_1)}{R_a}
\end{cases}
\tag{42}
$$

It should be noted that the first (propagation) equation in (42) is reduced to (35), i.e. to the equation for propagation in a MM with dipole-like MAs. However, dynamics of $\tilde{\rho}_{12,c}$ in case (42) will be affected by the antisymmetric mode and, therefore even in case of very closely spaced nanowires (when retardation can be neglected, which mathematically is expressed by $k\to 0$), the propagation equation will have its own peculiarities.

**Loss compensation by completely uncoupled QS ($\delta=0$)**

Fig. 8 shows propagation of the plane wave in (a) MMs with quadrupoles-like MAs only and (b) quadrupoles-like MAs and uncoupled QS. In the case of the quadrupole-like MAs there are three characteristic frequencies, namely the eigen frequency of a single nanoresonator (which remains the same as previously) $\omega_0=314THz$, symmetric and antisymmetric eigen frequencies of the quadrupole-like MAs (coupled nanoresonators) $\omega_s=324THz,\omega_a=304THz$ respectively. Stable propagation is achievable for both symmetric and antisymmetric eigenfrequencies, provided parameters for the full compensation (position of the gain maximum, concentration of the QS and pump level) are appropriately chosen. In Fig. 5 the stable propagation at the symmetric eigen frequency is shown; the stable propagation at the antisymmetric frequency looks the same and is not presented here.

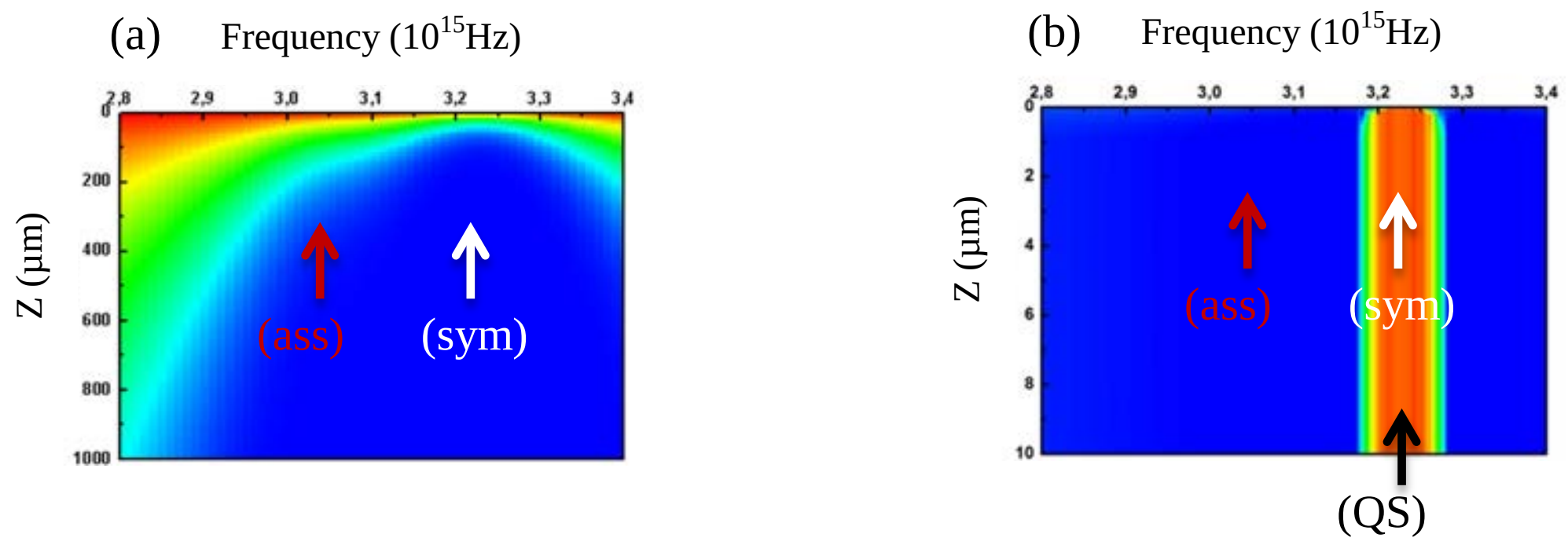

**Fig. 8: (a) Propagation of the plane wave through the MM consisting of quadrupole-like MAs. Two deeps correspond for antisimetryc 3.04 THz and symmetric 3.24 THz modes. Losses near symmetric mode are stronger than near antisymmetric. (b) Loss compensation regime with uncoupled QSs near the symmetric mode 3.24 THz. Positions of the symmetric, antisymmetric, and QS gain peak frequencies are shown by arrows.**

It has to be emphasized that all optical properties of the MM (like dielectric and magnetic response and consequently possibility of the negative refraction) are affected differently: an imaginary part of the effective dielectric constant in the stationary state is zero, while effective magnetic constant remains unaffected. It means in turn, that the loss compensation with the uncoupled QS enhances the effect of negative refraction; from the other side, number of the QSs necessary for the full compensation in the case of uncoupled QSs is pretty high and requires special technological methods.

**Loss compensation by completely coupled QS ($\delta = 1$)**

For the case of fully coupled QS the physical picture becomes (in compare with Fig. 8) even more reach and complicated. Due to the saturation caused nonlinearity the modes become coupled and energy transfer between the modes takes place. Moreover, there are three potential variants of the central gain positioning: one can place the QS gain center coinciding with the eigen nanoresonator frequency 3.14 THz as it was assumed for dipole-like MAs (see Figs. 6 and 7), and one can match the QS peak gain frequency with the symmetric or antisymmetric eigen frequencies. All three options have been investigated and the results are presented in Fig. 9. It has been found, that the resulted frequency pattern (Fig. 9) becomes extremely complicated and does not guarantee the stable propagation at the initially found eigen frequencies for symmetric, antisymmetric, and single nanoresonator oscillations. From the other side, stable propagation is possible and the frequency position of the stable propagation depends on the coupling efficiency.

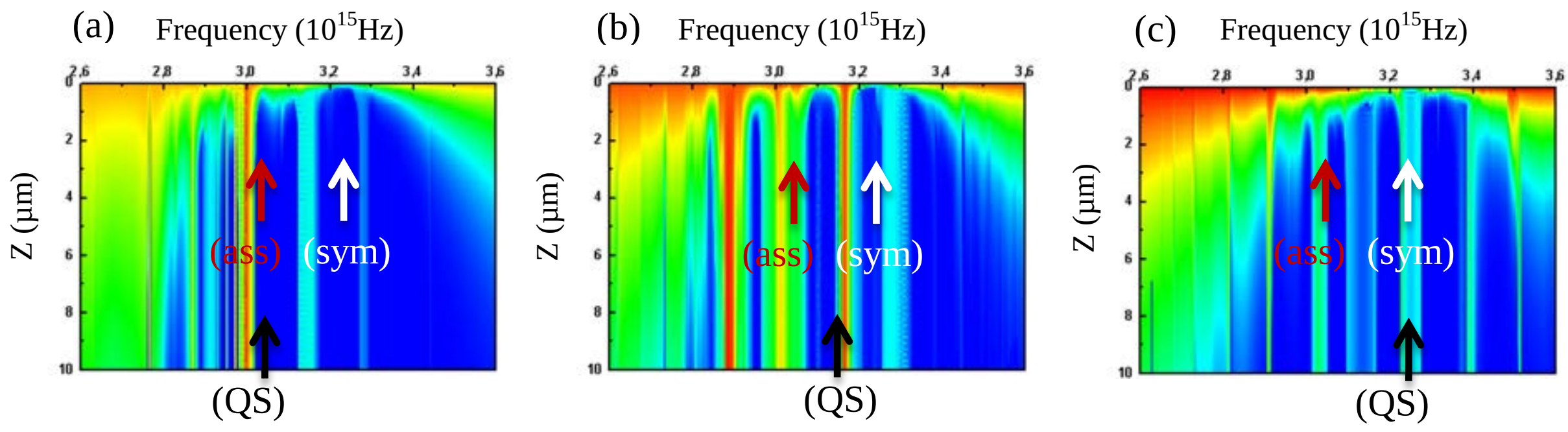

**Fig. 9: Propagation of the plane wave in the case of totally coupled QS in MM with quadrupole-like MAs. The amount of QS is fixed, $\omega_{QS}$ is varied (a) 3.04 x$10^{15}$Hz, (b) 3.14 x$10^{15}$Hz, and (c) 3.24 x$10^{15}$Hz respectively, $N_0 = 0.97$. There are certain frequencies (pattern depends on positioning of the QS peak gain) where losses are totally compensated and stabile propagation is possible. Positions of the symmetric, antisymmetric, and QS gain peak frequencies are shown by arrows.**

In real experimental realization the coupling efficiency is expected to be spatially inhomogeneous, and the resulted pattern is supposed to be even more complicated. In spite of the fact, that the positions of the stably propagating frequencies seem to be hardly predictable, the coupling (as it was mentioned for the dipole-like MAs) leads to significant reduction of the number of QS necessary for the full loss compensation, and from this point of view remains preferable way of the QS positioning in MM.

## 8. Dynamics of symmetric and antisymmetric modes in MAs at the propagation in case of completely coupled QS ($\delta = 1$)

The magnetic response of the MM differs MM from any natural material, and it would be extremely interesting to keep this property at the loss compensation scenario as well. In the frame of the developed in this work approach, the magnetic response depends on the magnitude of the antisymmetric oscillation mode – both magnetic and quadrupole moments are proportional to the magnitude of the antisymmetric mode. Here the magnitudes of the both antisymmetric $m_a$ and symmetric $m_s$ modes and their relation $m_a / m_s$ have been extracted and plotted in Fig. 10. It is seen, that clear domination of the antisymmetric mode (and consequently maximum magnetic response) could be obtained for the QS with the peak gain frequency around eigen frequency of a single nanoresonator. The results presented in Fig. 10 indicate that the energy of the inverted QS is transferred to the MAs and is redistributed between the symmetric and antisymmetric modes. Basically, the results reveal complex dynamics of the two nonlinearly coupled modes, where one of the modes is supported by a propagating plane wave. It has to be emphasized, that in this case possible nonlinear interaction between several plane waves was not considered: according to the accepted here SVA approximation a propagation of the external monochromatic wave was assumed. Experimentally it corresponds to the short propagation distances, where generation of new harmonics due to the nonlinear interaction is not significant; this operation mode is expected in the case of single or even multilayer MMs, which can be produced using the state of the art technologies. Nevertheless, in the case of simultaneous propagation of several waves with different carrier frequencies this effect has to be taken into account.

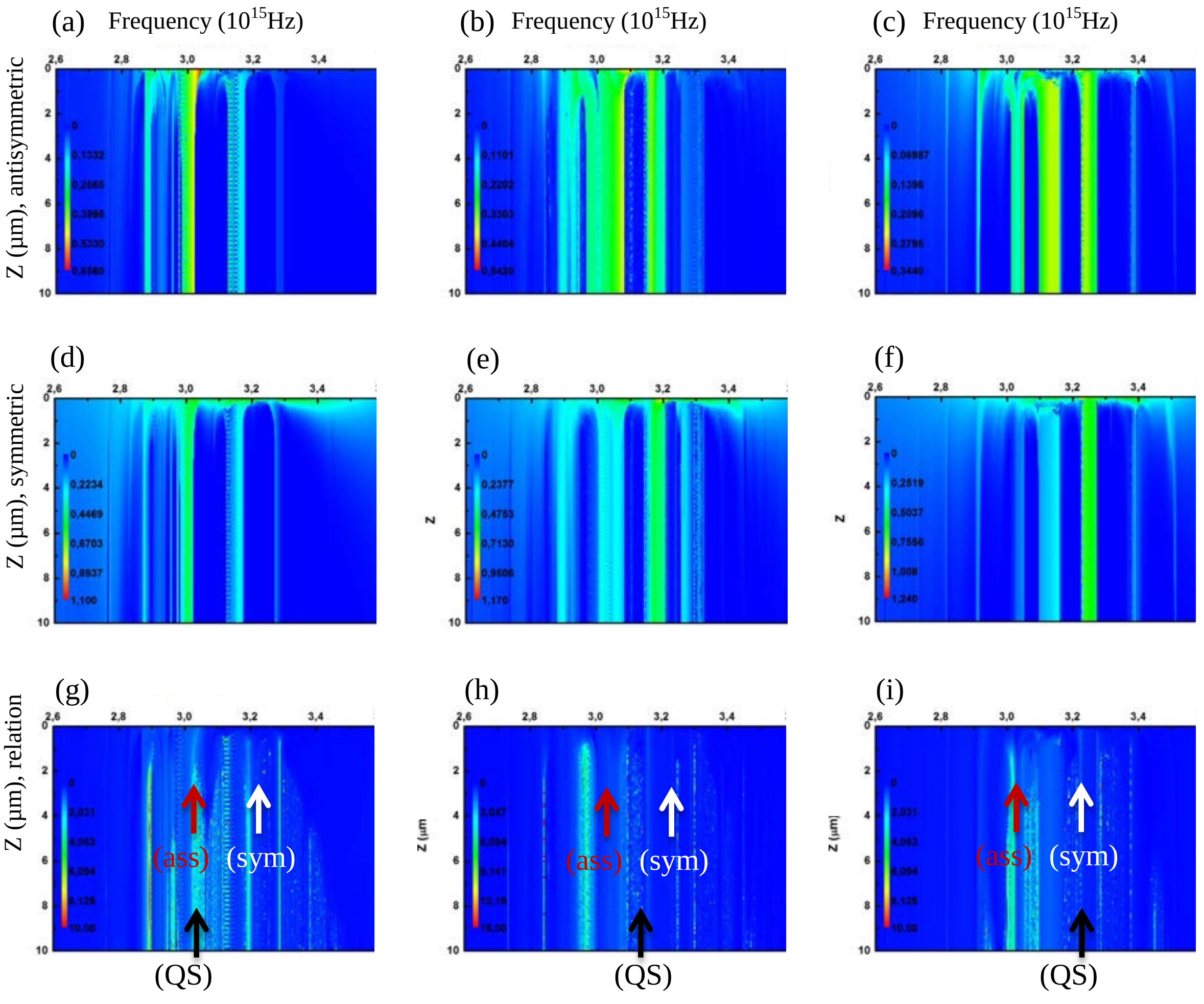

**Fig. 10: Dynamics of the amplitude of the (a, b, c) antisymmetric $m_a$ , (d, e, f) symmetric $m_s$ modes, and (g, h, i) their relation $m_a / m_s$ as a function of frequency for the same parameters as shown in Fig. 12.4: propagation of the plane wave in the case of totally coupled QS in MM with quadrupole-like MAs. The amount of QD is fixed, $\omega_{QS}$ is varied (a, d, g) 3.04x10$^{15}$Hz (eigen frequency of the symmetric mode of the quadrupole-like MAs), (b, e, h) 3.14 x10$^{15}$Hz (eigen frequency of the eigen mode of the single dipole-like MAs), and (c, f, i) 3.24 x10$^{15}$Hz (eigen frequency of the antisymmetric mode of the MAs), $N_0 = 0.97$. Positions of the symmetric, antisymmetric, and QS gain peak frequencies are shown by arrows.**

It is expected, that this kind of parametric interaction could be used for loss compensation as well; from the purely theoretical point of view, this parametric nonlinear interaction between modes is also interesting for the investigation of possible generation of the multifrequency stable states (colored stable states) which could probably exist in MMs. Otherwise, according to the results presented in Fig. 10 the propagation of a monochromatic plane wave is obviously unstable: the energy of the initial monochromatic mode will be redistributed over many frequencies as the plane wave propagates in the MM.

## 9. Conclusion

In conclusion, the problem of monochromatic plane wave propagation in the MM consisting of quantum MAs has been considered. The system of equations describing the plane wave propagation in the MM with the quantum MAs has been elaborated. Using the results about the existence of stable and unstable generation modes of the MA, the stable propagation has been observed for the case of the fully uncoupled and fully coupled QSs. It has been found, that for the same set of parameters the uncoupled QSs cannot compensate for the optical losses in the MM, which proves that the coupling between the plasmonic nanoresonators and QSs significantly increases the loss compensation efficiency. The case of the plane wave propagation in the MM has been considered only for the monochromatic plane wave propagation and hence requires further investigation.

The results presented in this paper resolve the discussion excited by [12] followed by [35-38]. The discussion was about basically two different systems: from one side, in [12] the system with fully coupled QS ($\delta = 1$ in terms of our paper, corresponding to the MM consisting of spasers) was considered, while from the other side the system "MM+gain" ($0 < \delta < 1$ in terms of our paper, corresponding to the MM consisting of MA plus just partially coupled QS) was discussed in [35-38]. In [4], where extensive numerical calculations were performed, the results indicate that the coupling between MA and active media was actually minimal, which of course should not cause any problems – the charge dynamics in MA remained the same and hence optical properties of the respective MM remained also almost unchanged excepting loss compensation. The physical realization in [4] is obviously much closer to the technologically achievable [5] and consequently closer for the real experimental conditions. From the other side, spased dynamical properties open up new intriguing features which otherwise unreachable with just MM consisting of uncoupled MA and quantum emitters.